\documentclass[fleqn,usenatbib]{mnras}

\usepackage{newtxtext,newtxmath}

\usepackage[T1]{fontenc}
\usepackage{ae,aecompl}
\usepackage{bm}		

\usepackage{graphicx}	
\usepackage{amsmath}	
\usepackage{amssymb}	

\usepackage{hyperref}
\usepackage{nicefrac}
\usepackage{listings}

\hypersetup{
  colorlinks   = true, 
  urlcolor     = blue, 
  linkcolor    = blue, 
  citecolor   = blue 
}
\def\kms{\hbox{km s$^{-1}$}}
\def\myr{\hbox{$M_{\odot}$ yr$^{-1}$}}
\def\msun{\hbox{$M_{\odot}$}}
\def\mdot{\hbox{$\dot{M}$}}
\def\teff{$\bm{T}_\text{eff}$}

\newcommand{\caps}[1]{{\scshape{#1}}}

\title[Space Density of Post--Period Minimum CVs]{The Space Density of Post--Period Minimum\\Cataclysmic Variables}

\author[Hern\'andez Santisteban et al.]{J. V. Hern\'andez Santisteban$^{1,2}$\thanks{E-mail:
j.v.hernandez@uva.nl (JVHS)},  C.
Knigge$^{1}$,  M. L. Pretorius$^{3,4}$, M. Sullivan$^{1}$ and \newauthor{B. Warner$^5$}\\
$^{1}$School of Physics \& Astronomy, University of Southampton, Southampton SO17 1BJ, UK\\
$^{2}$Anton Pannekoek Institute for Astronomy, University of Amsterdam, Science Park 904, NL-1098 XH Amsterdam, the Netherlands\\
$^{3}$South African Astronomical Observatory, PO Box 9, Observatory 7935, Cape Town, South Africa\\
$^{4}$Department of Physics, University of Oxford, Denys Wilkinson Building, Keble Road, OX1 3RH, Oxford, UK\\
$^{5}$Department of Astronomy, University of Cape Town, Rondebosch 7700, South Africa}

\date{Accepted XXX. Received YYY; in original form ZZZ}

\pubyear{2017}

\begin{document}
\label{firstpage}
\pagerange{\pageref{firstpage}--\pageref{lastpage}}
\maketitle

\begin{abstract}
Binary evolution theory predicts that accreting white dwarfs with
sub-stellar companions dominate the Galactic population of
cataclysmic variables (CVs). In order to test these predictions, it is necessary to identify these systems, which may be difficult if the signatures of accretion 
become too weak to be detected. The only chance
to identify such "dead" CVs is by exploiting their close binary nature. 
We have therefore searched the Sloan Digital Sky Survey (SDSS) Stripe
82 area for apparently isolated white dwarfs that undergo eclipses
by a dark companion. We found no such eclipses in either the SDSS or
Palomar Transient Factory data sets among our sample of 2264
photometrically selected white dwarf candidates within Stripe
82. This null result allows us to set a firm upper limit on the
space density, $\rho_0$, of dead CVs. In order to determine this
limit, we have used Monte-Carlo simulations to fold our selection
criteria through a simple model of the Galactic CV distribution. Assuming a $T_{WD}=7,500$ K, the
resulting 2$\sigma$ limit on the space density of dead CVs is $\rho_0
\lesssim 2 \times 10^{-5}$~pc$^{-3}$, where $T_{WD}$ is the typical
effective temperature of the white dwarf in such systems.
\end{abstract}




\section{Introduction}
A key prediction in binary evolution theory of accreting white dwarfs (WD) states that the highly evolved systems should harbour a sub-stellar or brown dwarf as donors \citep{Howell:1997aa,Kolb:1999aa,Knigge:2011aa}. These systems, commonly known as cataclysmic variables (CVs) accrete by Roche-lobe overflow of the secondary often via an accretion disc \citep[see reviews by][]{Warner:1995aa,Hellier:2001fr}. Their secular evolution is regulated by the removal of angular momentum of the system which shortens the orbital period $P_{orb}$ of the system \citep{Rappaport:1983ys}. As the donor continues to lose mass, the initial star will eventually transition to a sub-stellar regime. This critical point in the evolution roughly coincides with a reversal in the direction of the system's orbital period evolution \citep{Warner:1995aa}. Therefore, the properties of the donor at any given orbital period can serve as a proxy for the evolutionary state of any particular object \citep{Knigge:2006kx,Knigge:2011aa}. This is particularly useful since the orbital period derivative is typically too small to observe. 

The identification of a sub-stellar donor is therefore a good indicator that a particular system has already undergone the so-called period bounce \citep{Knigge:2011aa}. Still, establishing the sub-stellar nature of a particular donor star unambiguously is also difficult, since the secondary is hard to detect against the background of the optically brighter WD and accretion disc. It is no surprise that there is only one firm direct detection of a sub-stellar secondary \citep[SDSS J1433+1011,][]{Littlefair:2013aa,Hernandez-Santisteban:2016ab} among the $\sim20$ known period bounce candidates \citep{Littlefair:2008lr,Aviles:2010lr,Savoury:2011aa,Patterson:2011aa}.

The late evolution and final fate of these systems is still rather uncertain. As the binary separation increases, and the donor loses mass, the secular mass accretion rate is expected to drop to $\dot M \lesssim 10^{-12}-10^{-11}$ \myr\@, making it difficult to observing them unless they are in an unusual bright state, e.g., after a \textit{rare} nova or dwarf nova eruption. Establishing the existence and size of this population of period bouncers has been a major challenge for theorists and observers alike, mainly due to observational bias~\citep[e.g.][]{Pretorius:2007aa}. The recent discovery of hundreds of short orbital period systems has alleviated some of the problems \citep{Gansicke:2009aa}, and deep imaging and long-term monitoring campaigns have allowed us to find long-recurrence-time dwarf nova outbursts associated with short-period CVs \citep{Breedt:2014aa}. Nevertheless, it remains difficult to reconcile the observationally inferred numbers of post-period-minimum CVs with theoretical predictions. In fact, binary population synthesis studies predict that a full $\sim 40\% - 70$\% of all CVs should have reached this point of their evolution \citep{Kolb:1999aa,Goliasch:2015aa}.

CVs that have evolved well past the period minimum might be almost indistinguishable from isolated WDs, i.e. they may (appear to) be \textit{dead CVs}. In particular, they may not exhibit any of the classic accretion signatures that are commonly used to identify CVs: emission lines, a blue accretion disk continuum, short-time-scale variability (flickering). The only way to identify such systems is therefore to exploit their close binary nature: WD-dominated period bouncers are expected to have orbital periods in the range $75~{\mathrm min} \lesssim P_{orb} \lesssim 120~{\min}$ and dark secondaries with $R_2 \simeq 0.1~R_{\odot}$. These systems are therefore characterized by $R_2 \sim 0.1 a_{bin}$, where $R_2$ is the radius of the secondary and $a_{bin}$ the binary separation. The probability of such a system exhibiting an eclipse is then a substantial $P \simeq \sin{R2/a} \simeq 0.1$. Thus, if there is a significant population of such systems, $\simeq 10\%$ of them should be detectable via eclipses.

This technique has already been used in related areas. For example, searches for planetary eclipses in large samples of WDs were carried out by \citet{Fulton:2014aa} and \citet{Faedi:2011aa}, who were able to constrain the fraction of close Jupiter-size objects around WDs to be $\lesssim0.5$\% and $\lesssim0.3$\%, respectively. Similarly, 
\citet{Drake:2010aa} searched for eclipses in the light curves of $\simeq 12,000$ photometrically and spectroscopically selected WDs in the The Catalina Real-Time Survey (CRTS). They found 20 such systems with late M star secondaries, and 3 with possible sub-stellar companions \citep{Drake:2010aa}. Finally, \citet{Parsons:2013aa,Parsons:2015aa} searched for eclipsing systems among photometrically or spectroscopically identified WD binaries with main-sequence companions.

However, essentially all previous time-domain searches were insensitive to true "drop-outs" -- i.e. systems that effectively become invisible during eclipse. This is a key limitation: given that the WD completely dominates the light of the system, but $R_{wd} << R_2$, almost all eclipses of such systems will, in fact, be total. Here, $R_{wd}$ is the radius of the WD. Thus, during eclipses, such systems will typically disappear below the detection limit. The standard data product provided by existing time-domain surveys do not include sufficient information to identify such drop-outs. 

Here, we therefore present a new systematic study of 2264 photometrically selected WD in order to find transiting sub-stellar companions. We use data from the Sloan Digital Sky Survey \citep[SDSS,][]{York:2000aa} and the Palomar Transient Factory \cite[PTF,][]{Law:2009aa} spanning over $\sim$5 years and use custom-designed procedures to identify eclipses, including drop-outs. We then carry out a Monte-Carlo simulation of the Galactic CV population to convert our results into a constraint on the space density of dead CVs.

\section{The White Dwarf Sample}
\label{sec:wdsample}

Our analysis is focused on the SDSS \textit{Stripe 82} fields, a 270 deg$^2$ band centred on the celestial equator along the southern Galactic cap ($-50^{\circ}$~$\leq$~RA~$\leq$~$59^{\circ}$, -1.25$^{\circ}$ $\leq$ DEC $\leq$ 1.25$^{\circ}$). 
Sky positions located inside this band were typically scanned $\simeq 70-90$ times during the imaging survey. The SDSS data base therefore contains extensive variability information for sources within this region \cite[][see Section~3.1]{Sesar:2007aa,Bramich:2008aa,Bhatti:2010aa}. We further supplemented this data set with matching time-domain observations obtained by the PTF survey in Stripe~82 \cite[][see Section~3.2]{Law:2009aa}.

We constructed our Stripe 82 sample from the high signal-to-noise photometric catalogue of \citet[see Appendix \ref{sec:app1} for the \caps{SQL} query]{Annis:2014aa}, which is $>$95\% complete down to $g=21$. We identified  WD candidates from this catalogue via the Catalog Archive Server (CAS)\footnote{\url{http://cas.sdss.org/stripe82}} by applying the colour selection suggested by \citet{Girven:2011aa}, which they find to be 95\% complete in SDSS DR7.
This selection yielded 2264 WD candidates. Given the $62\%$ efficiency of the colour selection \cite{Girven:2011aa}, we estimate that our sample includes $\gtrsim$1600 bona-fide WDs. Most of the contaminants in our sample are quasars, which can mimic cool WDs in the SDSS colour space.
\begin{figure}
\begin{center}
\includegraphics[trim=0.0cm 0.8cm 0.3cm 0.0cm, clip, width=\columnwidth]{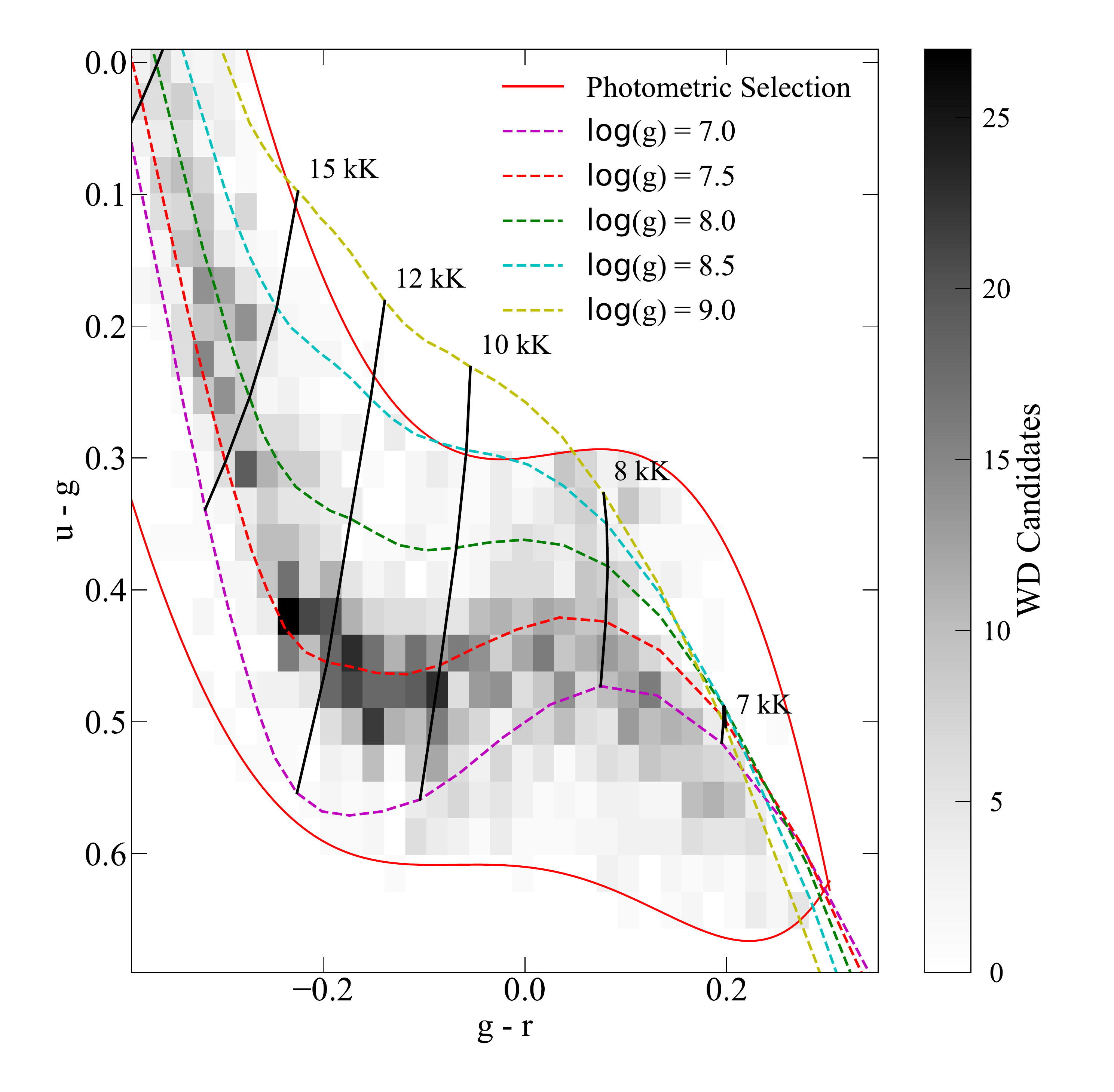}
\caption{Photometric selection and colours of the 2264 WD candidates in SDSS Stripe 82. Coulour-Coulour cuts are shown in red continuous line \citep{Girven:2011aa} and WD cooling tracks (dotted lines) are overlaid for different $\log (g)$ in 0.5 steps \citep{Holberg:2006aa}. The grey scale histogram shows the distribution of the WD candidates.}
\label{fig:photo_selection}
\end{center}
\end{figure}

\begin{figure}
\begin{center}
\includegraphics[trim=0.0cm 0.0cm 1.1cm 0.7cm, clip, width=\columnwidth]{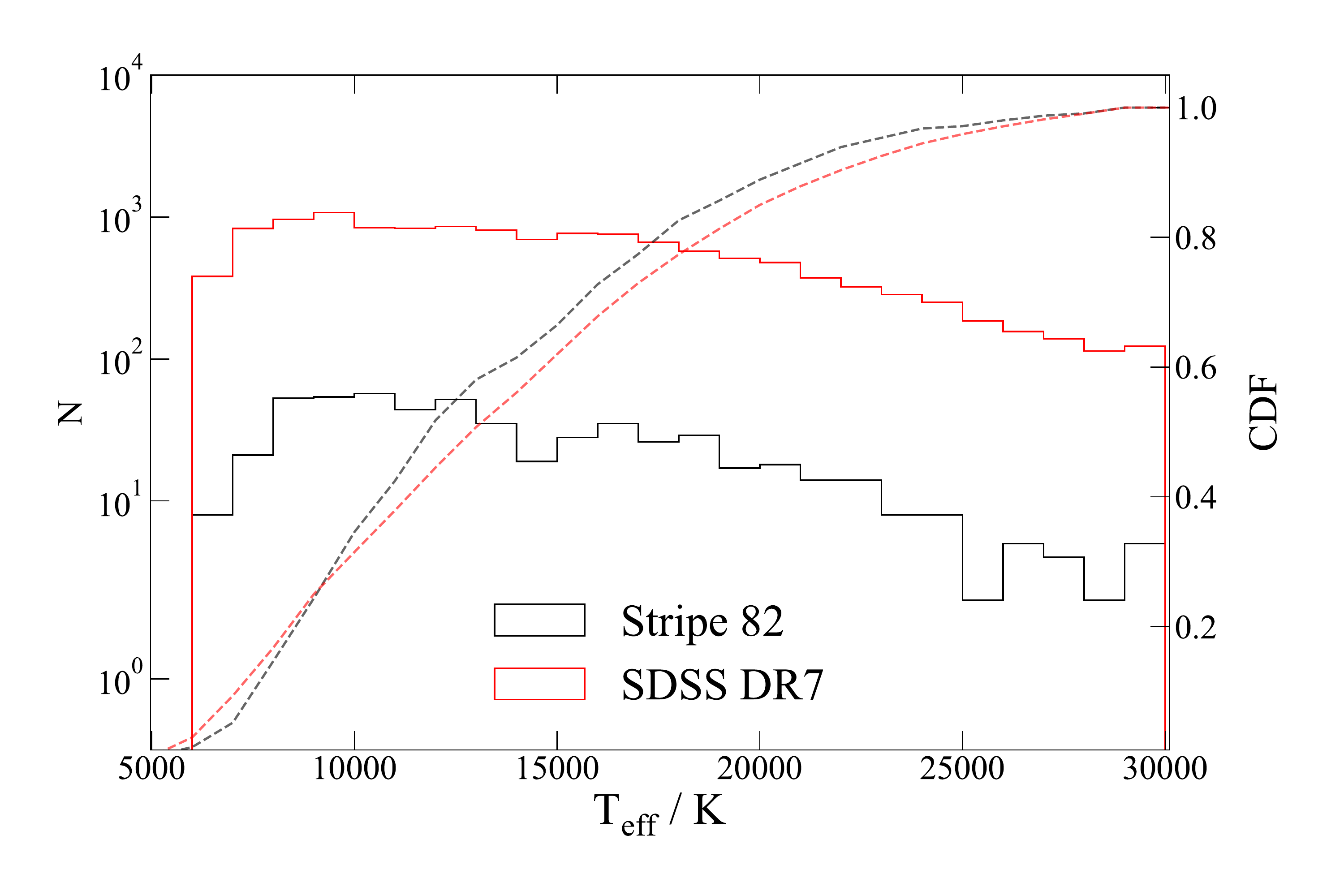}
\caption{Histograms and CDFs of the WD effective temperature obtained through spectroscopic modelling of \citet{Kleinman:2013aa}, for the full sample (\textit{red lines}) and the cross-match sample of our WD candidates (\textit{black lines}). }
\label{fig:temp_wds}
\end{center}
\end{figure}

The effective temperature, \teff, of the accreting WDs we are targeting is set by compressional heating in the freshly accreted envelope \citep{Townsley:2003aa}. As a consequence of their low accretion rates, \mdot\@, the WDs in period bouncers are expected to be quite cool, T$_{\textrm{\scriptsize eff}} \simeq 8,000 - 12,000$ K.
It is therefore critical that our WD selection method should not be biased against such low-temperature objects. Fig.~\ref{fig:photo_selection} shows the WD candidates with \cite{Holberg:2006aa} cooling tracks overlaid\footnote{\url{http://www.astro.umontreal.ca/~bergeron/Cooling Models}}. It is encouraging that most of our targets lie between 7,000 - 15,000 K. As an independent check, we cross-matched our candidate selection with spectroscopically confirmed WDs \citep{Kleinman:2013aa} and obtained a match for $\simeq 25$\% of our object selection. In Fig.~\ref{fig:temp_wds}, we show the distribution of \teff\@ for the full sample and our WD candidates. Our WD candidates have a similar distribution as the full sample with over half of them with  T$_{\textrm{\scriptsize eff}}\lesssim$ 15,000 K, confirming our sensitivity to sufficiently low-temperature WDs.

\section{Drop-Out and Eclipse Search}

We extracted light curves for our candidate WDs from the multi-epoch data obtained by both the SDSS and the PTF in Stripe 82. In this section, we will outline the procedures we used to search for eclipsing WDs in each of these surveys.

\subsection{SDSS sample}

\begin{figure}
\begin{center}
\includegraphics[trim=0.75cm .0cm 0.0cm 0.2cm, clip, width=\columnwidth]{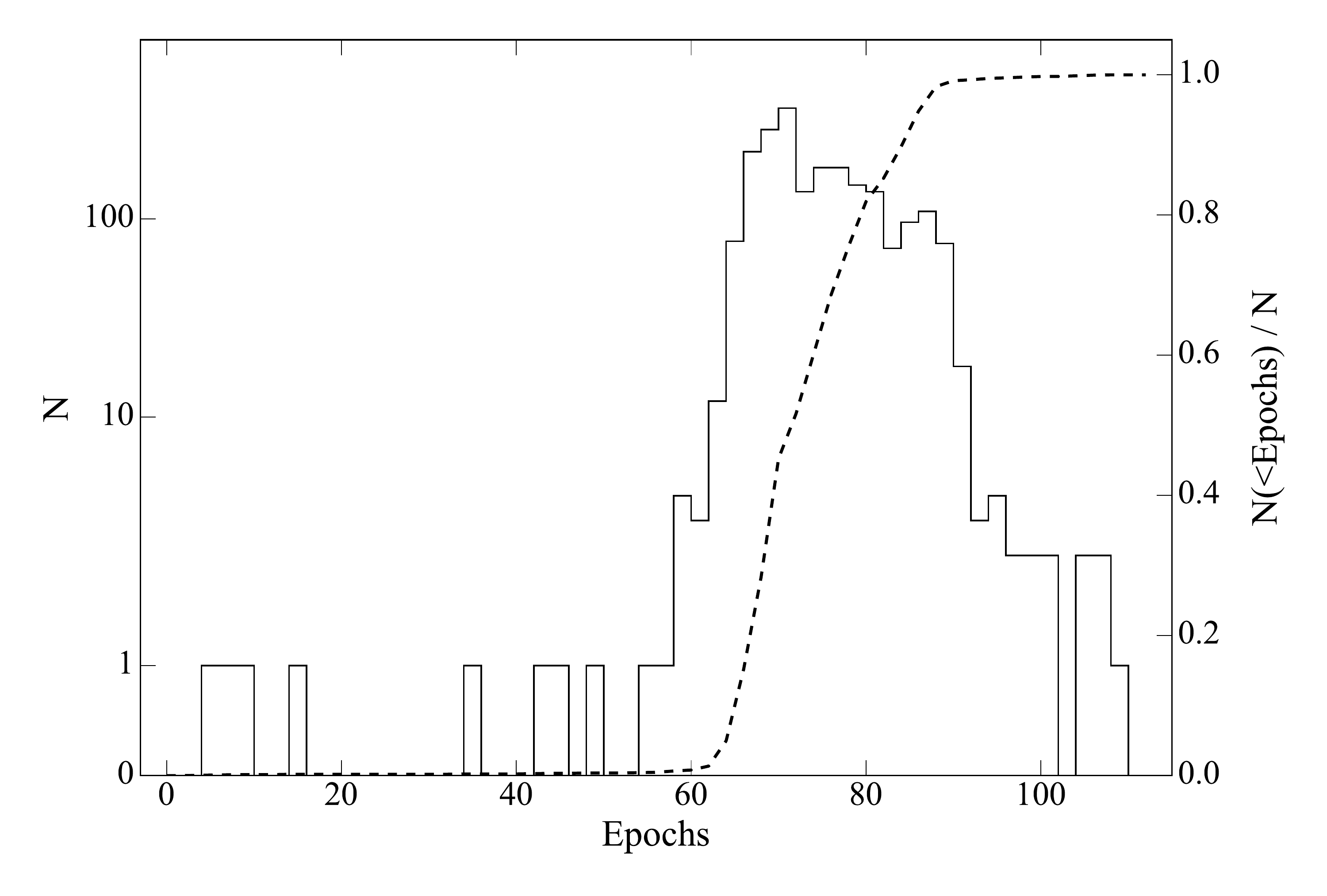}
\caption{Epoch distribution for the WD candidates in Stripe 82 given by SDSS. The cumulative distribution is plotted as dotted lines.}
\label{fig:sdd_distribution}
\end{center}
\end{figure}

The SDSS Stripe 82 dataset has already been extensively searched for variable objects \citep[e.g.][]{Sesar:2007aa,Bramich:2008aa,Bhatti:2010aa}. However, none of the earlier s studies included a drop-out search, i.e. none were sensitive to systems exhibiting sharp and deep eclipses, during which they fall below the detectability threshold of the survey (Ivezic, private communication). In order to search for eclipsing WDs -- including possible drop-outs -- we therefore proceeded as follows. We first queried the SDSS CAS database for all the available photometry (in a 3" radius) of our WD candidates selected in Section \ref{sec:wdsample}. The original survey observed each object $\sim40-100$ times in 302 observation runs of the Stripe 82 field, as shown in Fig.~\ref{fig:sdd_distribution}. In a similar way, we queried all fields where each individual WD should be contained. For every field observed at each particular location, we should expect a photometric measure i.e. a one-to-one relation between the two sets for all non-eclipsing systems. In practice, this picture is far from perfect since observational conditions and systematics of the data reduction pipeline contribute to false positives. In order to locate and discriminate potential eclipsing systems, we performed two basic tests. 

First, we discarded any observation where the object was localised near the edges of the CCD image. We do not expect to have a preferential bias towards finding eclipses near image edges. The remaining candidate were checked to rule-out false-positives caused by observational conditions such as cloud, bad seeing and/or cosmic rays. Specifically, for every candidate non-detection, we queried the specific frame for other objects that had similar brightness to the WD candidate (within $\Delta m\pm0.3$ mag) {\em in the co-added catalogue}. If these matched-brightness sources are also missing in the frame in question, the WD non-detection in this frame was considered to be a false positive, and it was excluded from the final sample. Following this step, only 271 frames were left for visual inspection. No true drop-outs were found among these. 
\subsection{PTF sample}
\label{sec:ptf}

\begin{figure}
\begin{center}
\includegraphics[trim=0.8cm 1.1cm 0.0cm 0.2cm, clip, width=\columnwidth]{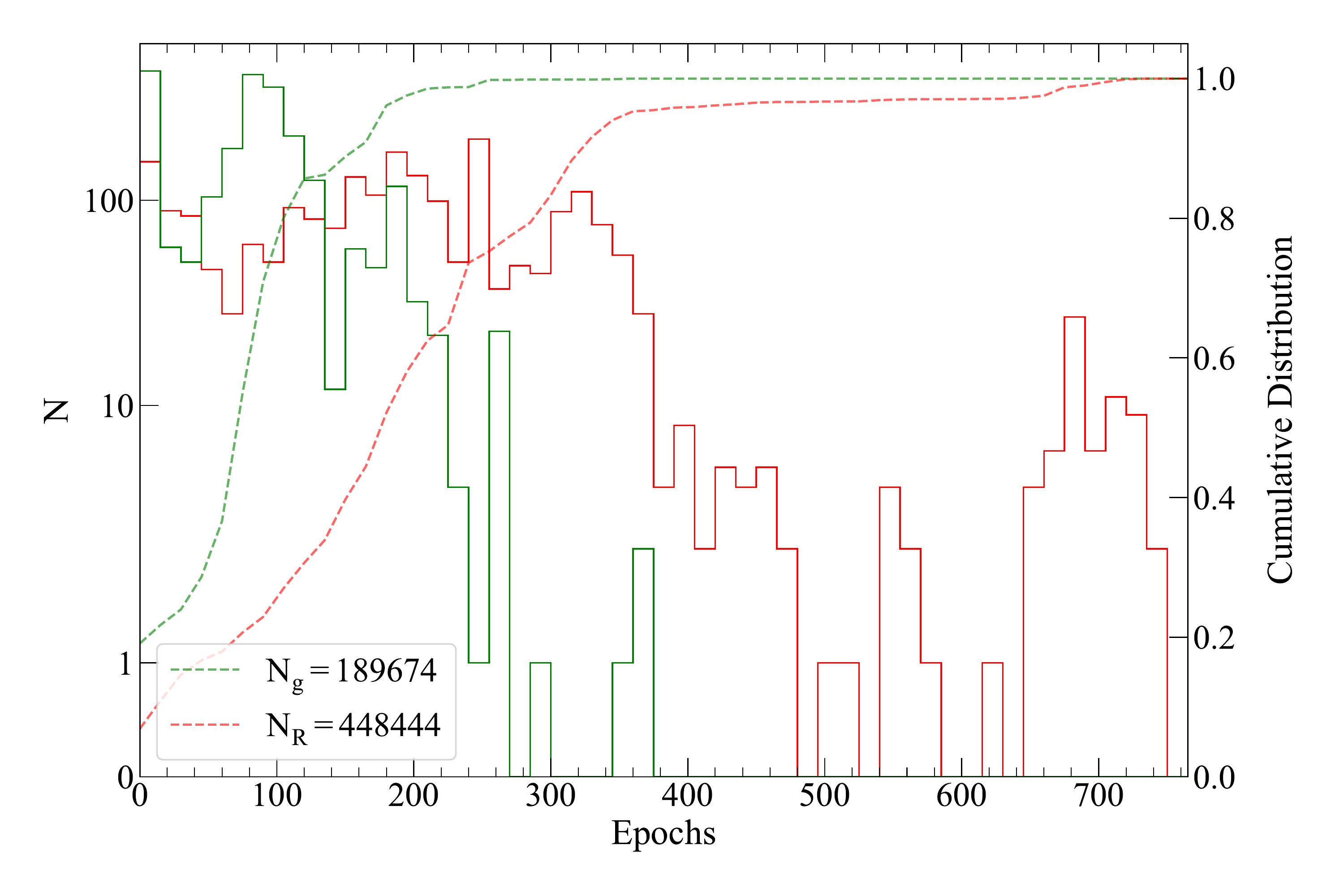}

\caption{Epoch distribution for the WD candidates in PTF in $R$ (red) and $g'$ (green) filters. The cumulative distributions are plotted in dotted lines for both filters.}
\label{fig:ptf_distribution}
\end{center}
\end{figure}

The Stripe 82 region was observed many times as part of the PTF survey in two filters (Mould $R$ and $g'$) over five years \citep[2009-2014,][]{Laher:2014aa}.
The light curves of all WD candidates were extracted using the pipeline described in \citet{Firth:2015aa}, which has been used extensively on PTF data. The pipeline performs point-spread-function (PSF) photometry and forced the extraction of each target in all the available fields. This allowed us to recover a flux estimate even when the object approached the limiting magnitude in a particular epoch. In total, our PTF sample contained 638,118 individual measurements of the 2264 photometrically selected WD candidates. The number of epochs available for a given object is highly variable, depending on the specific location of the object  in Stripe 82 (see Fig.~\ref{fig:ptf_distribution}). The number of epochs, $N$, for each filter was different. On average, a typical object was observed $N_R\sim200$ times in $R$ and $N_g\sim100$ times in $g'$.

\begin{figure*}
\begin{center}
\includegraphics[trim=0.0cm .0cm 0.2cm 0.0cm, clip, width=16cm]{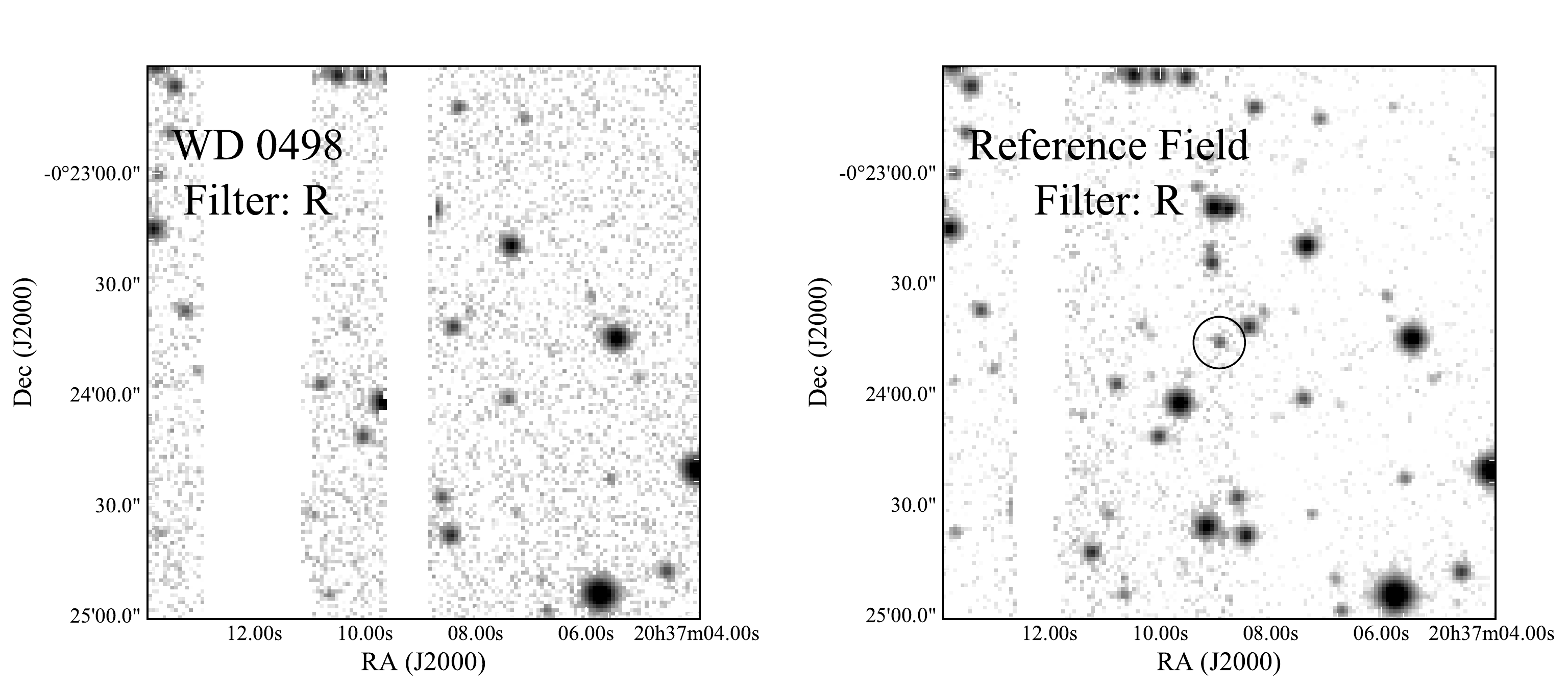}
\includegraphics[trim=0.0cm .0cm 0.2cm 0.0cm, clip, width=16cm]{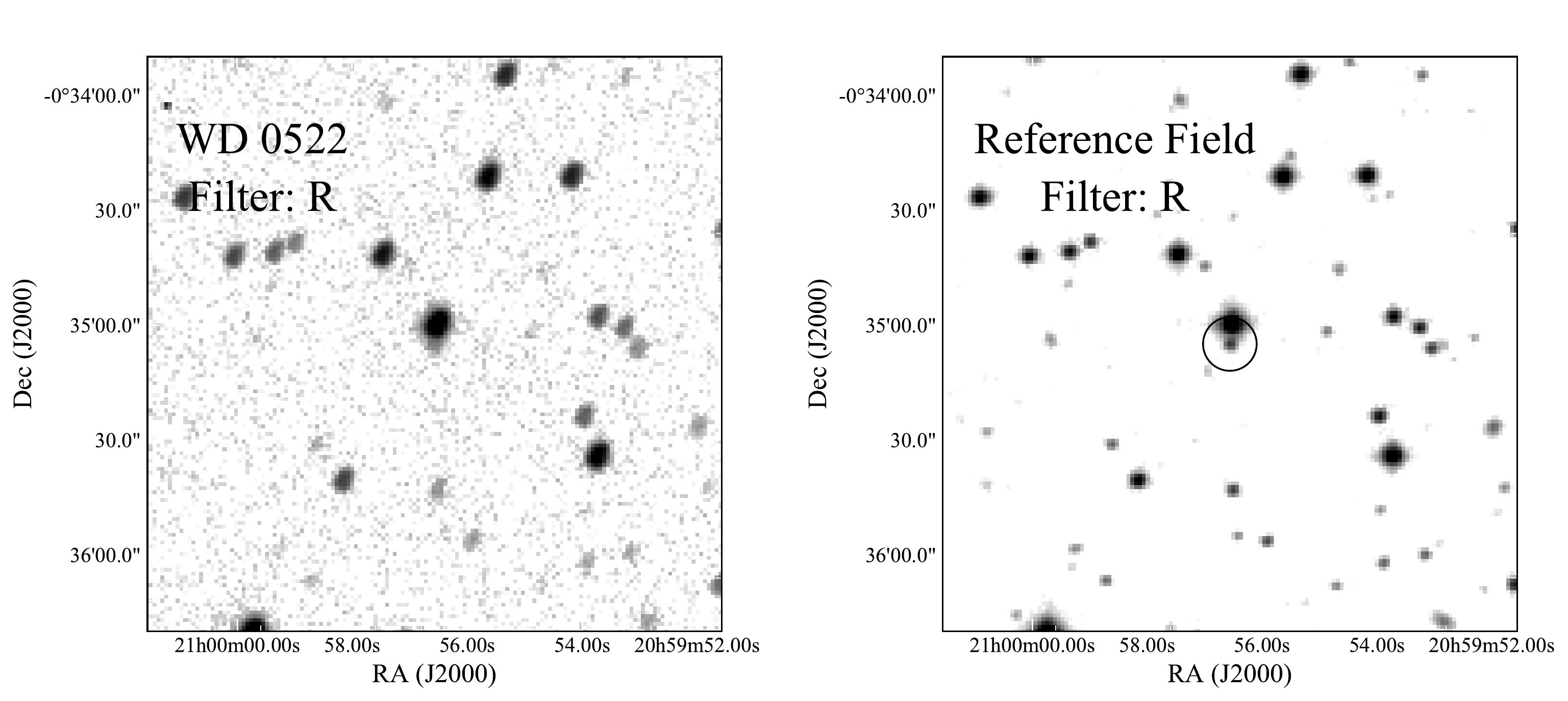}
\caption{Manual inspection of false positives in the PTF sample for two example WD candidates (WD 0498 in the top panel and WD 0522 in the bottom panel). \textit{Left:} The observed epoch for a WD candidate where the decision tree selected a possible non-detection. \textit{Right:} Reference image for the same field which clearly contains the WD. For WD 0498, the object is clearly missing due to its position in a missing column of the CCD. For WD 0522, the atmospheric conditions forced the extraction of the bright and nearby source producing a false positive.}
\label{fig:false_positive}
\end{center}
\end{figure*}

\begin{figure} 
\begin{center}
\includegraphics[trim=0.2cm 0.1cm 0.2cm 0.2cm, clip, width=\columnwidth]{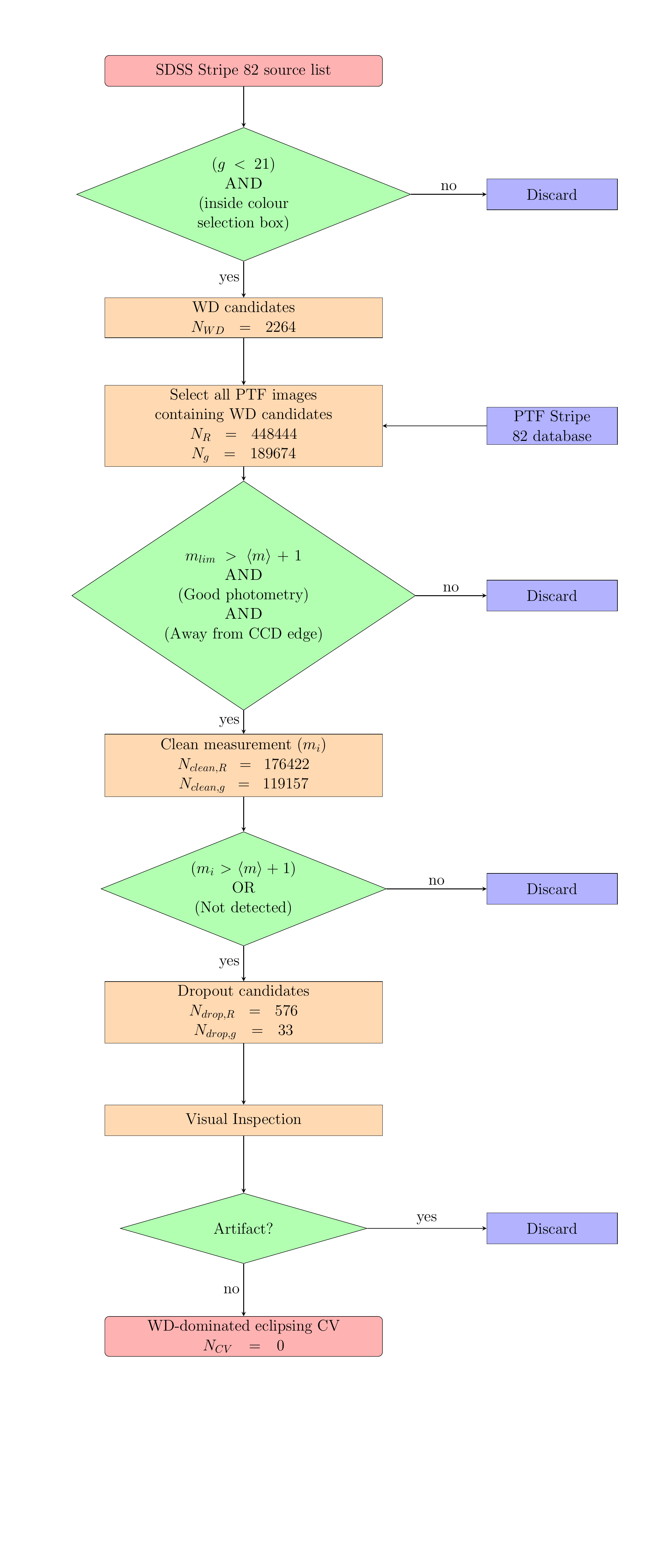}
\caption{Decision Tree for eclipse selection in the PTF data of Stripe 82.}
\label{fig:flowchart}
\end{center}
\end{figure}

In order to identify eclipses and drop-outs among our candidate WDs, we developed a decision tree, shown in Fig.~\ref{fig:flowchart}. This was designed to remove most false-positives in an unbiased way. We refer to this figure for the number epochs discarded in each step of the decision tree. 
First,  we again excluded all observations close to the edge of the detector.
In addition, we restricted attention to data points that were considered as \textit{good} measurements or obtained in good atmospheric conditions.
We also only considered candidate WDs whose average brightness in the PTF data, $\langle m_i \rangle$, was at least 1 mag brighter than the local background, $m_{lim} > \langle m_i \rangle +1 $, where $m_{lim}$ = 3$\sigma$ detection above the background. This allowed us to be confident that, in case of an eclipse, the signal to noise in that specific frame would be enough to discern a true dropout. Then, after this initial selection, we flagged an eclipse candidate as any observation where the pipeline could not retrieve a flux value ($m_i < m_{lim}$ ) or an observed flux drop greater than one magnitude was observed, $\Delta m = m_i - \langle m_i \rangle \geq 1$~mag. As a result, we obtained 606 candidates which were manually inspected. The most common cases were are exemplified in Fig.~\ref{fig:false_positive}. In the top panel, the dropout candidate was not observed due to its localisation on a faulty section of the detector. The bottom panel represents the most common case, where a WD candidate lies close to a brighter source. In bad atmospheric conditions, the WD would drop just below the detection limit and the pipeline would extract the bright near star. Thus, creating an artificial flux drop when analysing the overall mean flux. After the manual inspection, we found no true drop-outs nor eclipses in the sample.

\section{Limits on Space Density}

Given the null result of eclipsing or drop-out systems in our sample, we can estimate an upper limit on the space density of any hidden population of period bouncers. To this end, we performed Monte Carlo simulations in which we distributed fake samples of "dead CVs" across a simple model of the Milky Way and checked how many of them should have been recovered by our searches, for a given space density. In the following section, we will describe the Galactic model we used, the simulations we performed and the resulting constraints of the space density of dead CVs.

\subsection{The Galactic CV Model}

Our Galaxy model has been used previously to estimate CV space densities \citep{Pretorius:2007ab,Pretorius:2012aa,Pretorius:2013aa,Pretorius:2015aa}. We therefore only provide a brief description here.
\label{sec:galactic_model}
We model the space density of CVs, $\rho(z)$, as a function of their vertical position in the Galaxy as
\begin{equation}
\rho(z) = \rho_0 e^{-|z|/h},
\end{equation}
where $\rho_0$ is the Galactic plane density, $z$ is the distance from the Galactic plane, and $h$, is the characteristic scale height of the thin disc of the Galaxy. This assumption is reasonable since CVs have been shown to reside mostly in the Galactic disc \citep{Stehle:1997aa,Ritter:2003aa}. Radial gradients in the space density are ignored, since, for period bouncers with $T_{WD} \lesssim 12,500$~K, we are only sensitive to systems within $\simeq 1$~kpc. However, the simulations were computed for two different characteristic scale heights, $h=250$ pc (appropriate to a relatively young population) and $h=$450 pc (appropriate for older systems; this is more likely to apply to period bouncers). 

The density of neutral hydrogen also depends on $z$ in our model. The interstellar extinction is then found by integrating the column density along the line of sight for every WD in our sample. The density dependence of the interstellar medium (ISM) on $z$ is described by
\begin{equation}
\rho_{\textrm{\scriptsize ISM}} = \rho_{\textrm{\scriptsize ISM},0} e^{-|z|/h },
\end{equation}%
where $h_{ISM}=140$ pc and the ISM density in the mid-plane, $\rho_{ISM,0}$, was estimated by assuming $A_V=0.7$ mag kpc$^{-1}$ for $b=0^{\circ}$, in agreement with \citep{Amores:2005aa}.

\subsection{Eclipsing Systems}
\label{sec:eclipsing}
To determine the probability of a system to be eclipsing, we rely on the assumption that the WD is the dominant source in both $R$ and $g'$ bandpasses. Sub-stellar donors have low temperatures ($\lesssim 2500$ K) with spectral types later than $\textrm{SpT}\gtrsim\textrm{L1}$ \citep{Littlefair:2008lr,Hernandez-Santisteban:2016ab}. Their contribution to the optical is therefore negligible. Consequently, any eclipse will have a high contrast and can be modelled as a bright object being eclipsed by a dark, opaque and approximately spherical body. The half-width of such an eclipse in orbital phase units, $\varphi_{\nicefrac{1}{2}}$, is given by \citep{Dhillon:1991aa}
\begin{equation}
\label{eq:width}
\cos^2 (2\pi\varphi_{\nicefrac{1}{2}}) = \left[1-\left(\frac{R_2}{a}\right)^2\right]\sin^{-2} i,
\end{equation}
where $R_2$ is the radius of the donor, $a$ is the binary separation, and 
$i$ the inclination of the system. For Roche-lobe filling donors, $R_2/a$ is solely function of the mass ratio \citep{Eggleton:1983aa},
\begin{equation}
\label{eq:r2}
\frac{R_2}{a} = \frac{0.49 q^{2/3}}{0.6q^{2/3} + \ln\left(1+q^{1/3}\right)}\,\,\,\, ,\,\,\,\,\,\,\,\,\,\,\,\,\, 0<q<\infty,
\end{equation}
where $q = M_2/M_1$ is the mass ratio of the system. Combining Eq. \ref{eq:width} and \ref{eq:r2} yields the eclipse width solely as a function of $q$ and $i$. We can then  finally estimate the probability of a system to be observed in eclipse, $P_{\textrm{\scriptsize eclipse}}$:
\begin{equation}
P_{\textrm{\scriptsize eclipse}} = \frac{2\varphi_{\nicefrac{1}{2}}P_{orb}+t_{eff}}{P_{orb}}
\end{equation}
where $t_{eff}$ is the effective integration time (see Appendix \ref{sec:app_derivation}).

The probability of observing no eclipses, $P_{\textrm{\scriptsize no eclipse}}$, in a given light curve is:
\begin{equation}
P_{\textrm{\scriptsize no eclipse}} = (1-P_{\textrm{\scriptsize eclipse}})^N
\end{equation}
where $N$ is the number of observations of a WD candidate, given by our selection criteria described in Section~\ref{sec:ptf}. Then, the probability for each WD to have been observed with \textit{at least one} eclipse, $P_1$, becomes:
\begin{equation}
P_{1} = 1 - P_{\textrm{\scriptsize no eclipse}}
\end{equation}

The sum of the probabilities over every object in our mock sample then provides the expectation value for the number of eclipses we should observe in given mock population.

\subsection{Monte Carlo Simulations}

In order to predict the number of CVs in Stripe 82 as a function of space density, we performed a Monte Carlo simulation where we generated a Galactic sample of dead CVs in the Stripe 82 field. For computational efficiency, we created a grid of simulated dead CVs for four WD effective temperatures: 5000 K, 7500 K, 10000 K and 12500 K; and for two WD masses: M$_1 = 0.8$ M$_{\odot}$ \citep[mean mass for CVs,][]{Zorotovic:2011aa} and M$_1 = 0.6$ M$_{\odot}$ \citep[mean mass for field WDs,][]{Kleinman:2013aa}. The mass of the donor and orbital period at a given WD temperature, were taken from the revised evolutionary track of \citet[][extrapolating towards longer orbital periods and cooler WD temperatures]{Knigge:2011aa}. The luminosity of each system was assumed to be dominated by its WD primary, which was calculated from WD cooling tracks \citep{Holberg:2006aa}. After applying interstellar extinction, we finally selected those WDs that complied with our selection criteria ($g<21$). We discuss the implications of these assumptions in Sec.~\ref{sec:caveats}

\begin{figure}
\begin{center}
\includegraphics[trim=0.9cm 1.1cm 0.3cm 0.2cm, clip, width=\columnwidth]{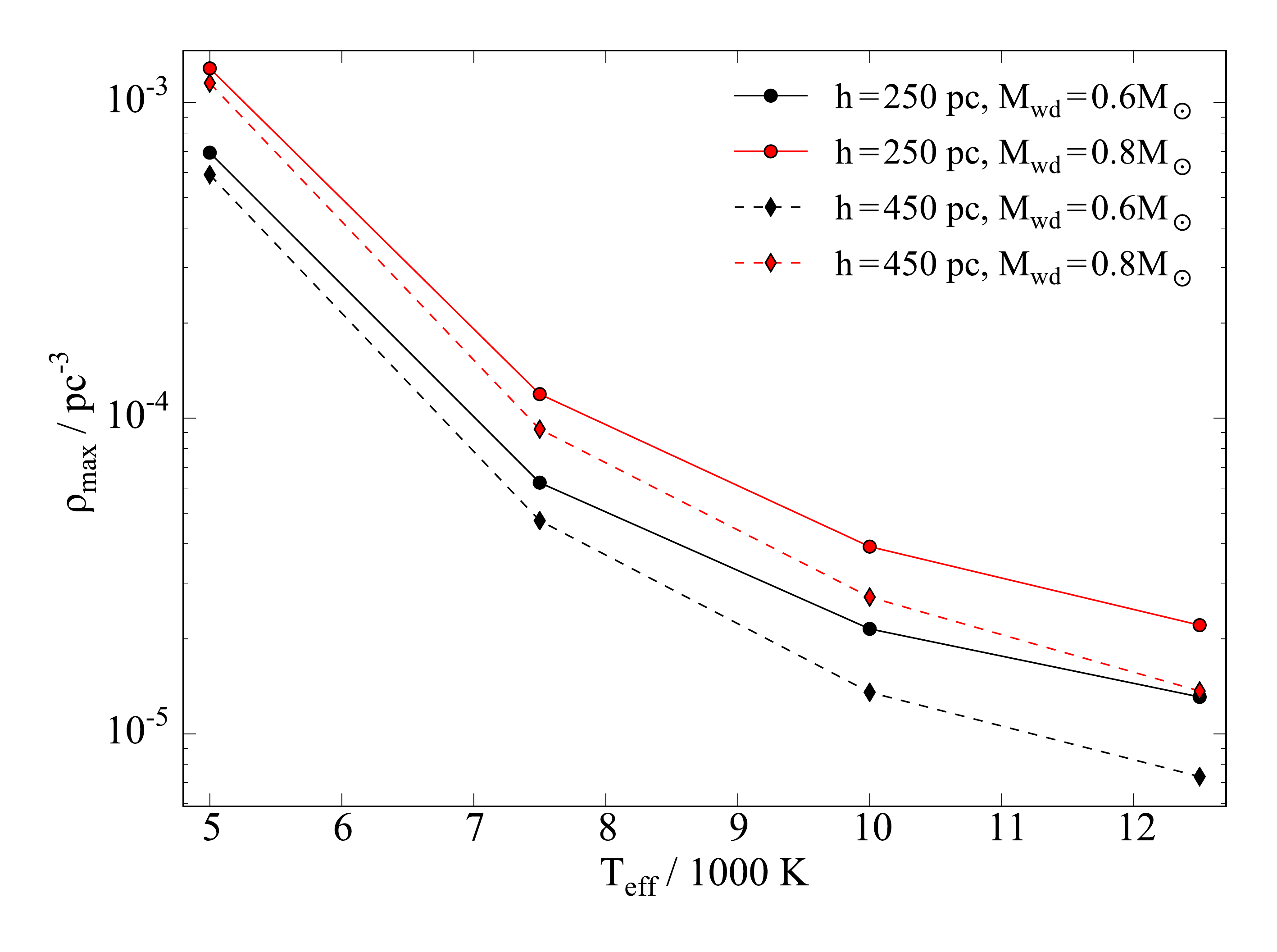}
\caption{Space density upper limit for post-period minimum CVs as a function the white dwarf effective temperature. We show the upper limits for different combinations of scale heights (250 and 450 pc) and white dwarf mass (0.6 and 0.8 \msun).}
\label{fig:space_density}
\end{center}
\end{figure}

Having generated a mock WD sample, we then 
calculated the number of epochs for every simulated target based on the observational constraints of the survey (see Sec.~\ref{sec:ptf}). Specifically, for each WD in a given mock sample, we found the nearest WD candidate in the real sample and assigning the same number of observable epochs to our simulated WD. We then estimated the probability of observing at least one eclipse for each simulated WD and, finally, obtained the expectation value for number of eclipsers or drop-outs we should have detected (see Section~\ref{sec:eclipsing}). These calculations are carried out for each mock sample, i.e. for each combination of $T_{WD}$ and $M_1$. To determine the upper limit on the $\rho_0$ associated with a given mock sample, we calculated the mid-plane space density for which we would have expected to find 3 eclipsers/drop-outs. A null detection is then a 2$\sigma$ result. These limits on the space density are shown in Fig.~\ref{fig:space_density} as a function of $T_{WD}$.

Our results provide firm upper limits on any hidden population of dead CVs. Since known systems near the period minimum have WD temperatures close to 12,500~K \citep[e.g. SDSS~J1433+1011][]{Hernandez-Santisteban:2016ab}, the WDs in period bouncers are likely to be cooler than this. The dependence of our limiting $\rho_0$ on $T_{WD}$ can be approximated as a power law for $7,500<T_{WD} <12,500$, yielding (for $h = 450$~pc)
\begin{equation}
\rho_0 \lesssim 1.8 \times 10^{-5}  \left(\frac{T_{WD}}{10^4~{\mathrm K}}\right)^{-4.5} {\mathrm{pc}}^{-3}.
\end{equation}
As may be expected, this limit depends significantly on the characteristic WD temperature (and thus the orbital period) of the hidden population. Cooler WD primaries -- corresponding to longer-$P_{orb}$, lower-$\dot{M}$ period bouncers -- are fainter and thus harder to recover. 

\subsection{Caveats}
\label{sec:caveats}

We explored the robustness of our results by assuming limiting cases for the system components, particularly the effects of $P_{orb}$ and $M_2$. In order to accommodate the viable parameter space of period bouncers, we re-ran the Monte-Carlo simulations with parameters appropriate for (i) systems close to the period minimum: $P_{orb}=1.3$ hr and $M_2=0.06$ \msun; and (ii) highly evolved CVs: $P_{orb}=3.0$ hr and $M_2=0.02$ \msun. Changing these parameters primarily results in a vertical shift at a given $T_{WD}$. For WD temperatures close to 10,000 K, we find maximum variations in the upper limit to be $\sim50$\%. For our maximally evolved case, $P_{orb}=3.0$ hr and $M_2=0.02$ \msun, $T_{WD} = 7,500$~K may be a more appropriate estimate. This parameter combination yields $\rho_0 \lesssim8\times10^{-5}$ pc$^{-3}$.

A possible caveat to the results presented in this study is the validity of the initial assumption that dead CVs resemble isolated WDs. There are other accreting systems which possess very low mass transfer rates $\dot{M}\lesssim10^{-12}$ \myr\@ such as AM CVn \citep{Solheim:2010aa} and even NS systems \citep[e.g.][]{Degenaar:2010aa}, that display features of accretion e.g. emission lines, blue continuum and outbursts. This fact would prevent us from selecting them accurately in deep multi-colour surveys. However, those systems would, of course, have been included in previous CV space density estimates. Our goal here is specifically to set limits on the size of any yet-to-be-discovered CV population.

\section{Discussion}

\begin{figure*}
\begin{center}
\includegraphics[trim=0.9cm .0cm 0.3cm 0.2cm, clip, width=13cm]{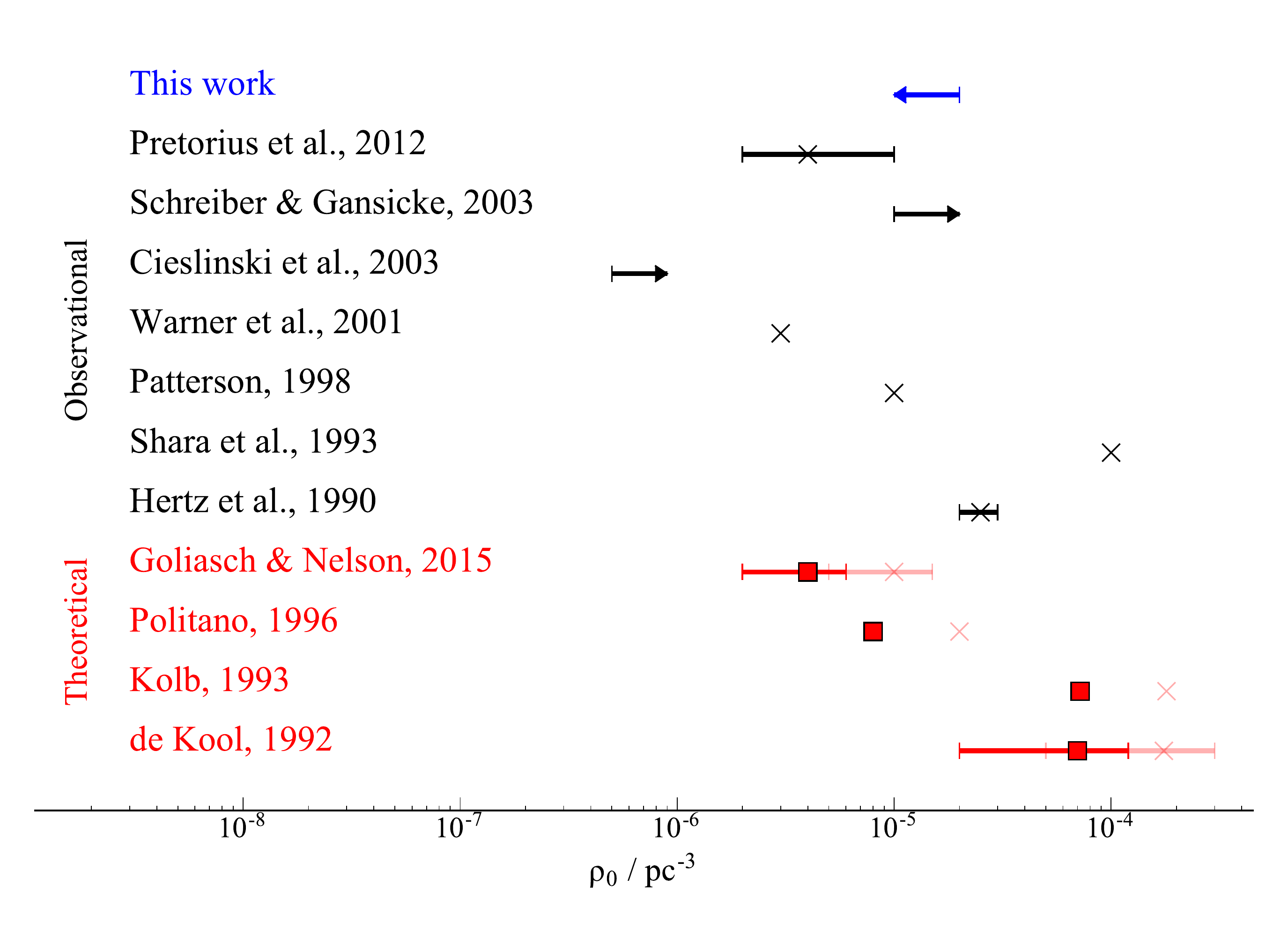}
\caption[Space density comparison for CVs]{Space density comparison for CVs. The total $\rho_0$ are shown in \textit{crosses} and the scaled \textit{dead CVs} space densities are shown in \textit{squares}. Theoretical estimates of the space density of CVs have been taken from \citet{de-Kool:1992aa,Kolb:1993ab,Politano:1996aa,Goliasch:2015aa} and observational constraints from \citet{Hertz:1990aa,Shara:1993aa,Patterson:1998aa,Warner:2001aa,Cieslinski:2003aa,Schreiber:2003aa,Pretorius:2012aa}.}
\label{fig:rho_summary}
\end{center}
\end{figure*}

Figure~\ref{fig:rho_summary} summarizes how our space density constraint compares to other observational estimates and theoretical predictions. It is important to note here that all previous observational estimates refer to the {\em total} population of CVs, subject to the assumption that all systems belonging to this population are in principle identifiable as such. By contrast, the work presented here has targeted the elusive population of period bouncers. More specifically, our goal was to test for the existence of a large population of "dead" CVs, i.e. WD-dominated, low-$\dot{M}$ systems with no obvious accretion signatures.

The theoretical predictions shown with crosses in Figure~\ref{fig:rho_summary} are also for the total CV population. In order to provide a rough point of comparison for our results, we also show an estimate of the period bouncer space density for each of these predictions, assuming that 40\% of CVs have evolved past the period minimum. This is a conservative estimate, since theoretical predictions of this fraction typically range from $\simeq 40\% - 70\%$.

Figure~\ref{fig:rho_summary} shows that recent theoretical and observational constraints are actually in reasonable agreement, both for the total CV population and for period bouncers. In particular, the CV space density predicted by \citep{Goliasch:2015aa} is in good agreement with the empirical estimate of \citep{Pretorius:2012aa}, as well as with the upper limit on the space density of dead CVs we have provided here.

How can we reconcile this agreement with the fraction of period bouncer candidates in observed CV samples being far smaller than 40\%? The easiest explanation is that selection effects are to blame \cite[see, for example,][]{Pretorius:2007aa}. Indeed, \citet{Goliasch:2015aa} find that with a selection function proportional to $\dot{M}$, the fraction of period bouncers is reduced to typically $\simeq 5\%$. The method we have developed here will soon be able to test this idea quantitatively. The next generation of large-scale synoptic surveys, e.g. the \textit{Large Synoptic Survey Telescope} \citep[LSST,][]{Ivezic:2008aa} and the \textit{Javalambre J-PAS} survey \citep{Benitez:2014aa}, will allow us search for eclipses and drop-outs among much larger WD samples covering huge areas of the sky. This should easily yield the roughly $\times 10$ improvement in sensitivity that will be necessary to test the latest theoretical models.

\section{Conclusions}

We have carried out a search for "dead" CVs, i.e. low-$\dot{M}$ WD-dominated systems with no observable accretion signatures. This was motivated by theoretical predictions that the overall CV population may be dominated by highly evolved, ultra-faint CVs that have already evolved far beyond the period minimum. Our goal was therefore to test for the presence of a large, previously undetected population of such period bouncers, hiding among apparently isolated WDs.

Dead CVs are expected to have dark, sub-stellar companions, thus eclipsing systems among them will show deep eclipses and would likely produce drop-outs in light curves produced by synoptic surveys. We therefore created a sample of candidate WDs in the well-observed SDSS \textit{Stripe 82} and retrieved all photometric data for this region from two surveys: SDSS and PTF. We developed a pipeline that searched for eclipses and drop-outs in both surveys, for all of our $\sim$2,600 WD candidates. In total, we analysed $\sim10^6$ individual measurements spanning over 5 years of observations. 

No eclipsing dead CVs were found. This null detection allowed us to constrain, for the first time, the space density of such systems. To this end, we carried out Monte Carlo simulation based on a simple Galactic model and the selection cuts used in the PTF sample. Based on this, we were able to derive an upper limit of $\rho_0 \lesssim 1.8 \times 10^{-5}  (T_{WD}/10^4~{\mathrm K})^{-4.5}$ pc$^{-3}$, where $T_{WD}$ is the characteristic WD temperature among any such hidden population. For period bouncers, it is expected that $8000~{\mathrm~K} \lesssim T_{WD} \lesssim 12,000~{\mathrm~K}$. Our limit is lower than several theoretical estimates in the literature, \citep[e.g.][]{de-Kool:1992aa, Kolb:1993aa}, but consistent with the most recent predictions \citep{Goliasch:2015aa,Schreiber:2016aa}.

\section*{Acknowledgments}
JVHS acknowledges support via studentships from CONACyT (Mexico) and the University of Southampton. JVHS is also supported by a Vidi grant awarded to N. Degenaar by the Netherlands Organization for Scientific Research (NWO). MS acknowledges support from EU/FP7-ERC grant No. [615929]. We would like to thank A. Drake and E. Breedt for useful discussion on the early stages of the project, as well as S. Prajs and C. Frohmaier for help on PTF data reduction. We would like to thank the anonymous referee for the suggestions that improved this work.\\
Funding for the SDSS and SDSS-II has been provided by the Alfred P. Sloan Foundation, the Participating Institutions, the National Science Foundation, the U.S. Department of Energy, the National Aeronautics and Space Administration, the Japanese Monbukagakusho, the Max Planck Society, and the Higher Education Funding Council for England. The SDSS Web Site is \url{http://www.sdss.org/}. The SDSS is managed by the Astrophysical Research Consortium for the Participating Institutions. The Participating Institutions are the American Museum of Natural History, Astrophysical Institute Potsdam, University of Basel, University of Cambridge, Case Western Reserve University, University of Chicago, Drexel University, Fermilab, the Institute for Advanced Study, the Japan Participation Group, Johns Hopkins University, the Joint Institute for Nuclear Astrophysics, the Kavli Institute for Particle Astrophysics and Cosmology, the Korean Scientist Group, the Chinese Academy of Sciences (LAMOST), Los Alamos National Laboratory, the Max-Planck-Institute for Astronomy (MPIA), the Max-Planck-Institute for Astrophysics (MPA), New Mexico State University, Ohio State University, University of Pittsburgh, University of Portsmouth, Princeton University, the United States Naval Observatory, and the University of Washington. This research made use of \caps{astropy}, a community-developed core Python package for Astronomy \citep{Astropy-Collaboration:2013aa}, \caps{matplotlib} \citep{Hunter:2007aa} and \caps{aplpy} \citep{Robitaille:2012aa}.




\bibliographystyle{mnras}
\bibliography{bibliography} 




\appendix
\section{SQL Query for WD candidates in Stripe 82 Coadd run}
\label{sec:app1}
SQL code to retrieve the WD candidates following the colour -coulour cuts in \citet{Girven:2011aa} and the best selection criteria for \citet{Annis:2014aa} for the coadd run of Stripe~82.
\lstset{language=SQL,breakatwhitespace=true,breaklines=true}
\begin{lstlisting}
SELECT objid,ra,dec,u,g,r,i,z,run,rerun, camcol, field,primtarget, psfMag_u, psfMag_g, psfMag_r, psfMag_i, psfMag_z, flags, psfmagerr_u, psfmagerr_g, psfmagerr_r, psfmagerr_i, psfmagerr_z
FROM PhotoObj
WHERE (run = 106 OR run = 206)
AND (u-g)>=-20.653*power(g-r,5) +10.816*power(g-r,4) +15.718*power(g-r,3) -1.294*power(g-r,2)-0.084*(g-r)+0.300
AND (u-g)<=-24.384*power(g-r,5) -19.000*power(g-r,4) +3.497*power(g-r,3) +1.193*power(g-r,2)+0.083*(g-r)+0.610 
AND (g-r)<=-0.693*power(r-i,2) +0.947*(r-i) +0.192 
AND (g-r)>=-1.320*power(r-i,3) +2.173*power(r-i,2) +2.452*(r-i)-0.070 
AND (r-i)>=-0.560 
AND (r-i)<=0.176*(i-z)+0.127 
AND (r-i)<=-0.754*(i-z)+0.110
AND ((flags & 0x10000000) != 0)
AND ((flags & 0x8100000c00a4) = 0)
AND (((flags & 0x400000000000) = 0)
OR (psfmagerr_r <= 0.2 AND psfmagerr_i<=0.2 AND psfmagerr_g<=0.2)) AND (((flags&0x100000000000) = 0) OR (flags & 0x1000) = 0)
AND type = 6
AND mode = 1
AND g<21
ORDER BY g
\end{lstlisting}
\bsp	

\section{White dwarf candidate list}
We present the photometric properties of the white dwarf selection for both SDSS and PTF in Table~\ref{tab:wd_selection}.
\begin{table*}
\centering
\begin{tabular}{ccc|cccc|cccc}
\hline
WD  & RA & Dec & SDSS $g'$&\multicolumn{3}{c}{PTF $g'$ }& SDSS $r'$ &\multicolumn{3}{c}{PTF $R$ } \\
id & deg & deg & mag & $\langle m\rangle$ & $\langle\sigma_m\rangle$ & r.m.s. & mag  & $\langle m\rangle$ & $\langle\sigma_m\rangle$ & r.m.s. \\
\hline
0000  & 309.16595 & -0.66203 & 18.17 & 18.13 & 0.02 & 0.10 & 18.017 & 17.98 & 0.02 & 0.049 \\
0001  & 309.16952 &  0.36834 & 17.06 & 17.08 & 0.01 & 0.05 & 17.407 & 17.51 & 0.02 & 0.035 \\
0002  & 309.29225 & -0.74514 & 18.08 &  $\cdots$ &  $\cdots$ &  $\cdots$ & 18.056 &  $\cdots$ &  $\cdots$ &   $\cdots$ \\
0003  & 309.38997 &  0.32968 & 18.67 & 18.71 & 0.03 & 0.17 & 18.963 & 19.10 & 0.05 & 0.092 \\
0004  & 309.49033 &  0.98303 & 16.23 & 16.36 & 0.03 & 0.07 & 16.509 & 16.62 & 0.01 & 0.058 \\
0005  & 309.72902 &  0.65325 & 18.09 & 18.12 & 0.02 & 0.03 & 18.417 & 18.52 & 0.03 & 0.074 \\
0006  & 309.72314 &  0.62537 & 18.66 & 18.66 & 0.02 & 0.05 & 18.475 & 18.49 & 0.03 & 0.155 \\
0007  & 309.64339 & -0.17908 & 18.74 & 18.81 & 0.03 & 0.06 & 19.062 & 19.23 & 0.06 & 0.098 \\
0008  & 309.59899 &  1.10977 & 17.93 & 17.92 & 0.02 & 0.09 & 18.154 & 18.26 & 0.03 & 0.093 \\
0009  & 309.78807 &  1.10119 & 16.40 & 16.43 & 0.01 & 0.08 & 16.713 & 16.80 & 0.01 & 0.019 \\
0010  & 309.96603 &  0.66858 & 17.44 & 17.55 & 0.01 & 0.04 & 17.812 & 18.02 & 0.02 & 0.046 \\
0011  & 310.45577 &  0.59884 & 18.76 & 18.83 & 0.03 & 0.09 & 19.100 & 19.22 & 0.05 & 0.072 \\
0012  & 310.53853 & -0.39504 & 18.69 & 18.74 & 0.03 & 0.08 & 19.121 & 19.23 & 0.05 & 0.110 \\
0013  & 310.49572 &  0.05706 & 18.78 & 18.87 & 0.03 & 0.09 & 19.321 & 19.45 & 0.06 & 0.083 \\
0014  & 310.56374 &  0.99690 & 18.93 & 18.95 & 0.03 & 0.10 & 19.143 & 19.26 & 0.05 & 0.074 \\
\multicolumn{11}{c}{$\vdots$ }\\
2556  & 55.53135 &  0.56991 & 20.92 & 20.13 & 0.12 & 0.43 & 20.849 & 20.49 & 0.18 & 0.380 \\
2557  & 55.49306 &  0.59192 & 20.82 & 21.09 & 0.24 & 0.37 & 21.069 & 20.94 & 0.44 & 0.089 \\
2558  & 55.72258 & -0.47331 & 20.89 &  $\cdots$ &  $\cdots$ &  $\cdots$ & 20.774 & 21.24 & 0.44 & 1.589 \\
2559  & 58.25896 & -0.58321 & 20.98 &  $\cdots$ &  $\cdots$ &  $\cdots$ & 20.980 & 21.07 & 0.31 & 0.735 \\
2560  & 59.45808 & -0.25499 & 20.72 &  $\cdots$ &  $\cdots$ &  $\cdots$ & 20.495 & 20.42 & 0.21 & 0.429 \\
\hline

\end{tabular} 
  \caption[Properties of the WD candidates]{Photometric Properties of the WD candidates such as the mean magnitude ($\langle m\rangle$), the mean of the 1$\sigma$ errors of all the photometry ($\langle\sigma_m\rangle$), and the root-mean-square (r.m.s.). WD entries were not on the PTF footprint and therefore no photometric information was recovered. The complete table is available in electronic format at \url{Vizier}.}
  \label{tab:wd_selection}
 \end{table*}
 
\section{Effective exposure time}
\label{sec:app_derivation}

The probability that an individual snapshot with an exposure time, $t_{exp}$, overlaps with an eclipse of duration, $t_{eclipse}$, is given by
\begin{equation}
P = \frac{t_{\textrm{eclipse}}+t_{\textrm{exp}}}{P_{orb}}\,\,\,\, .
\end{equation}
We can then estimate this probability since $t_{\textrm{eclipse}} = 2\pi\varphi_{\nicefrac{1}{2}}$ and $t_{exp}=60$ s for the PTF survey. However, the exposure time will be affected by the fact that the selection criteria (see Sec.~\ref{sec:eclipsing}) impose a flux drop of $\Delta m \geq1.0$. An observation where the eclipse is partially observed can therefore achieve the selection criteria, thus modifying the $t_{exp}$. We can then define an \textit{effective exposure time}, $t_{eff}$, given by
\begin{equation}
t_{\textrm{eff}} = t_{\textrm{exp}} + \Delta t\,\,\,\, ,
\end{equation}
where $\Delta t$ is a correction factor. In order to calculate $\Delta t$, we need to do some approximations for the eclipse. 
\begin{figure}
\begin{center}
\includegraphics[trim=0.3cm .0cm 0.3cm 0.2cm, clip, width=\columnwidth]{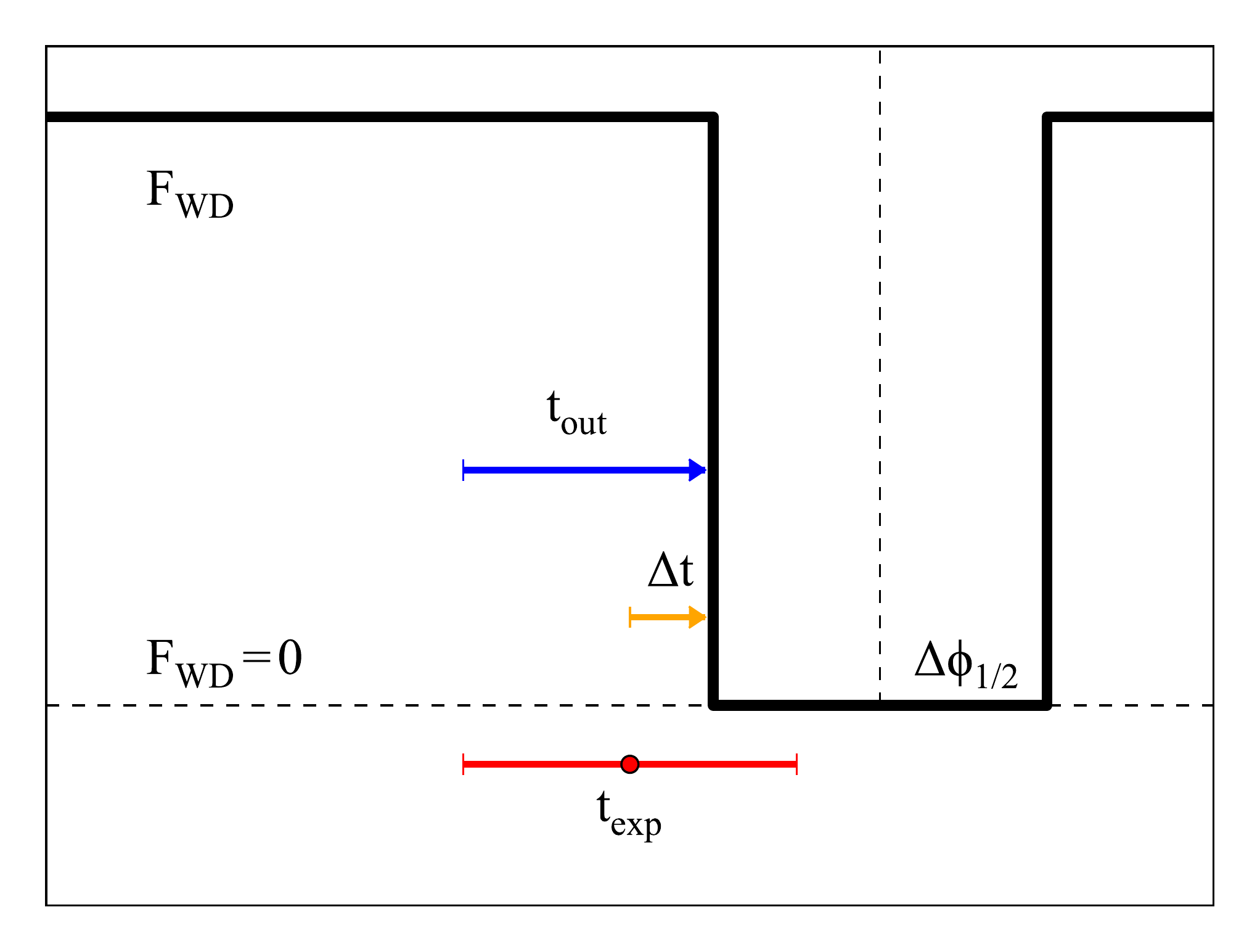}
\caption[Schematic of a WD eclipse and the determination of the effective exposure time]{Schematic of a WD eclipse and the determination of the effective exposure time. The vertical dashed line represents mid-eclipse and the horizontal dashed line, the zero-flux level. The exposure time ($t_{\textrm{exp}}$, red ), the time outside eclipse ($t_{\textrm{out}}$, blue) are marked for reference.}
\label{fig:eclipse_diagram}
\end{center}
\end{figure}
Due to the high-contrast between the WD and the sub-stellar donor contributions at optical wavelengths, the eclipses can be approximated as a light source eclipsed by an opaque body, as exemplified in Fig.~\ref{fig:eclipse_diagram}. Therefore, the flux, $f$, measured in a single exposure will be determined by the ratio of time spent outside eclipse, $t_{out}$, over the entire exposure time:
\begin{equation}\label{eq:b1}
f = \frac{t_{\textrm{out}}}{t_{\textrm{exp}}}\,\,\,\, .
\end{equation} This is valid as long as the WD is a constant source of light.

Then, we can obtain the amount of time the eclipse has to be inside an individual exposure to comply with a given $f$. From Fig.~\ref{fig:eclipse_diagram}, we can define $\Delta t$ as
\begin{equation}\label{eq:b2}
\Delta t = t_{\textrm{out}} - \frac{1}{2}t_{\textrm{exp}}\,\,\,\,.
\end{equation}
Due to the symmetry of the eclipse, the factor $\Delta t$ has to be accounted twice since it can happen before and after the eclipse. Therefore, by multiplying by a factor of 2 in Eq.~\ref{eq:b2} and  combining with Eq.~\ref{eq:b1} the final correction factor for the exposure time is:
\begin{equation}
\Delta t = (2f - 1)t_{\textrm{exp}}\,\,\,\,,
\end{equation}
where we have set the flux drop to $f=0.4$, the equivalent of $\Delta m=1$ as defined in the photometric selection.

\bsp	
\label{lastpage}
\end{document}